# Title: NIR-II Fluorescence Project Technology for Augmented Reality Surgical Navigation

**Authors:** Yuhuang Zhang,[1]† Xiaolong Liu,[2]† Zihang Liu,[1]† Chao Liu,[2] Jie Yang,[2] Jian Feng,[2] Siying Sun,[1] Zhe Feng,[1] Xiaoxiao Fan,[2] Hui Lin,[2] Jun Qian[1,2]*

**Affiliations:**

1 State Key Laboratory of Extreme Photonics and Instrumentations, Centre for Optical and Electromagnetic Research, College of Optical Science and Engineering, International Research Center for Advanced Photonics, Zhejiang University, Hangzhou 310058, China

2 Sir Run Run Shaw Hospital, School of Medicine, Zhejiang University, Hangzhou, 310016, China

† These authors contributed equally to this work
* **Corresponding author** Jun Qian qianjun@zju.edu.cn

**One Sentence Summary:** Projecting NIR-II fluorescence images onto the anatomical sites creates a real-time, augmented reality map to guide surgery.

**Abstract:** NIR-II fluorescence imaging provides superior tissue penetration and clarity, yet its clinical use in surgical navigation is hindered by a critical workflow issue. Surgeons must divert their attention between the operative field and external monitors, increasing cognitive load and disrupting procedures. Current strategies have failed to resolve this fundamental problem. Here, we developed a co-axial NIR-II fluorescence projection navigation system to enable real-time, in situ visualization. This system creates an intraoperative augmented reality by directly projecting high-precision, pseudocolored fluorescence images onto the surgical field, spatially integrating functional signals with patient anatomy. Validated through in vitro, in vivo, and clinical patient studies, our system eliminates visual field switching, reduces intraoperative distraction, and preserves natural stereoscopic vision. This approach represents a paradigm shift toward a more coherent, efficient, and ergonomically optimized optical imaging modality for surgical navigation.

## INTRODUCTION

Fluorescence imaging in the second near-infrared window (NIR-II, 900–1880 nm) has emerged as a promising technology for dynamic observation of living biological samples. Compared to the visible (380–760 nm) and first near-infrared (NIR-I, 700–900 nm) regions, NIR-II imaging offers enhanced tissue penetration and superior image clarity. Pioneering applications have significantly advanced the clinical translation of NIR-II fluorescence imaging for surgical navigation, as it enables precise intraoperative delineation of lesion boundaries, thereby improving diagnostic and surgical accuracy. Tian and his colleagues first extended NIR-II fluorescence imaging to guided surgery of liver tumor. Their research demonstrated that NIR-II fluorescence imaging surpassed NIR-I fluorescence imaging in superior tumor-to-normal-tissue ratio and enhanced tumor detection sensitivity(*1-4*). They further explored its clinical utility for precise resection of brain gliomas, renal tumors and cervical tumors. Our group has extended NIR-II imaging technology to hemodynamic monitoring in diabetic foot ulcers and perforator flap perfusion assessment, enabling real-time intraoperative vasculature visualization and providing high-precision quantitative data for hemodynamic decision-making during surgery(5). However, the aforementioned NIR-II fluorescence imaging remains constrained by its lack of anatomical context, displaying only pseudocolored signal on a screen. This compels surgeons to mentally integrate fluorescence images with the actual surgical field, relying heavily on experience for interpretation. To address this limitation, we have developed a series of dual-channel fusion imaging systems integrating NIR-II fluorescence with visible-light bright-field modalities(6, 7). These systems simultaneously delivered high-resolution anatomical structures (via visible-light bright-field) and lesion information (via NIR-II fluorescence), with precise spatial registration and overlay of both data streams, enhancing surgeons' perception and spatial targeting accuracy toward lesions(7). Collectively, these advancements establish an integrated technical network for NIR-II fluorescence-guided precision surgery, marking transformative progress toward clinical translation.

Despite the demonstrated clinical value of NIR-II fluorescence imaging for intraoperative navigation, one fact that cannot be ignored is that NIR-II fluorescence signal is invisible to the naked eye, making current display systems exhibit critical limitations in information transfer efficiency and surgical ergonomics. Existing approaches adopted a spatially segregated visualization paradigm: a dedicated monitor rendered pseudocolored NIR-II fluorescence data delineating pathological margins, while direct intraoperative observation provided anatomical context(8-10). Although dual-channel fusion imaging systems resolved the spatial segregation between pseudocolored NIR-II fluorescent signal and bright-field anatomical structures, they failed to eliminate the surgeon's frequent visual field switching during operation. Surgeons must still alternate their gaze between the display screen and the surgical site, a workflow that induced attention fragmentation, prolonged operative time, and elevated procedural risks. Consequently, stringent demands are placed on hand-eye-brain coordination, the learning curve for mastering the technique is substantially extended, and overall surgical complexity is increased*(11, 12)*. In view of this, a strategy named hybrid-fluorescence imaging surgical navigation was proposed by our team, utilizing NIR-II fluorescence for overall but large-depth pre-surgical lesion tracing, followed by precise surgical manipulation under direct vision using visible fluorescence signal(13). This technique partially reduced screen dependency of surgeons, shortened operative time, and decreased procedural complexity. However, it remains limitations such as the requirement to co-administration of two kinds of clinically-approved fluorophores, potential safety concerns associated with ultraviolet excitation for visible fluorescence, and most critically, the persistent need for a screen to observe the initial NIR-II fluorescent signal, failing to fundamentally resolve the issue of visual field switching. A brand-new paradigm is thus urgently needed.

To address the clinical demand for real-time intraoperative imaging and eliminate surgeon visual-field switching, we introduce a pioneering NIR-II fluorescence projection navigation system. This innovation achieves in-situ visualization by projecting pseudocolored NIR-II fluorescence images directly onto surgical tissues, enabling spatial fusion beyond traditional displays, as well as intraoperative augmented reality. By dynamically overlaying green pseudocolored fluorescent data with high precision, the system eliminates the need for gaze switching between the surgical site and external monitors, significantly reducing procedural distractions and interruptions to improve surgical coherence and efficiency. In addition, it enhances the three-dimensional depth perception of surgeons through natural stereoscopy. Through animal studies and clinical patient trials, we validated the exceptional performance of our NIR-II fluorescence projection navigation system. We believe this work will usher in a new era for optical imaging surgical navigation.

## RESULTS

### System Characterization and Performance Validation

A system for co-axial NIR-II fluorescence imaging and visible light projection was developed, and its architecture schematic is shown in Fig 1A. In operation, the system captures NIR-II fluorescence images from the anatomical site of biological samples, which are then processed and projected in real-time back onto the original anatomical site as a corresponding green light pattern. The core of the system is a customized dual-channel objective that functions as both the imaging objective for the NIR-II channel and the projection objective for the visible light channel.

Within this assembly, a dichroic element aligns the NIR-II imaging and visible projection paths to be co-axial in the object space, thereby avoiding parallax-induced position errors. The imaging and projection channels share a common front lens group, allowing for unified adjustment of the working distance. The optical design is optimized to minimize distortion differences between the NIR-II and visible channels, which is critical for reducing projection displacement. A green LED is selected to backlight the LCD module. This choice leverages the peak spectral sensitivity of the human eye, enabling sufficient perceptual brightness with minimal power consumption. The system is designed to overlay NIR-II fluorescence information onto the anatomical site in bright surgical environments. To ensure high contrast of the projected pattern against ambient illumination, the captured images are converted into near-binary patterns for projection. This approach prioritizes the clear visualization of structural boundaries and positional information over preserving a full grayscale range, ensuring that the image guidance is unambiguous to the operator. The system features a large working distance range of 24-100 cm, offering the flexibility to accommodate diverse applications, from fine surgical navigation to large-area examinations. We identified 28 cm as the optimal working distance, as it provides a balance between spatial resolution and the field of view, while also ensuring the system does not physically impede the surgeon. We conducted comprehensive experiments to quantify the performance of NIR-II fluorescence projection system, including its spatial resolution and positioning accuracy. Unlike standalone imaging or projection systems, the overall spatial resolution of our integrated system is jointly determined by the pixel sizes of the InGaAs camera and the LCD panel, and the optical resolution of the objective. To ensure the display was not the limiting factor, we selected an LCD panel with a pixel pitch (~12 μm) smaller than that of the InGaAs camera (~15 μm).

System resolution was evaluated by imaging and projecting a 1951 USAF resolution test chart. At a working distance of 28 cm, we first captured an image of the test chart under NIR illumination. The projection was then frozen, and the chart was replaced with a white paper screen to visualize the projected image. The details resolved in the projection reflected the end-to-end resolution of the system. As shown in the photograph of the projection (Fig 1B), the smallest resolvable element was Group 2, Element 3, corresponding to a system resolution of approximately 220 μm. At this working distance, the field of view was approximately 10 cm × 13 cm with a depth of field of 7 cm, providing sufficient space for surgical maneuvers without frequent refocusing. Notably, the projected pattern brightness reached 2600 lux, which is comparable to the ambient illumination in a surgical room outside the direct beam of a shadowless lamp, ensuring clear pattern visibility.

Next, we evaluated the system's positioning accuracy. In a similar experimental setup, we imaged a printed checkerboard pattern under visible and NIR illumination. The image was then frozen for projection, and the checkerboard was replaced by a white paper screen. An ancillary, fixed visible-light-sensitive camera captured images of both the original printed checkerboard and the resulting projected pattern. The displacement between the corresponding corner points in the two images was quantified as the system's positioning errors (Fig 1C).

With the objective focused at the 28 cm working distance, we measured the positioning errors at various distances within the depth of field (24-31 cm). The results, shown in Fig 1D, indicate that the positioning error varied minimally within this range, though it slightly increased with working distance (Fig 1E). Overall, the positioning error was $0.24 \pm 0.01$ mm across the 7 cm depth of field, providing both precise positioning and ample space for surgical tasks. At the optimal working distance of 28 cm, the positioning error was less than 0.15 mm. The system's high spatial resolution and low positioning error meet the sub-millimeter accuracy required for precision surgery.

### *In Vitro* Positioning with NIR-II Fluorescent Materials

We first evaluated the performance of NIR-II fluorescence projection system in positioning NIR-II fluorescent materials *in vitro*. Aqueous dispersions of Indocyanine Green (ICG), aggregation-induced emission (AIE) dots (TT3-oCB)(*14*), PbS/CdS quantum dots (QDs) and polymer dots (Pdots) (L1057) were tested(*15, 16*), with water used as a negative control. All tested fluorophores exhibit absorption near 800 nm and emission in the NIR-II window (Fig 2A). The dispersions were placed in centrifuge tubes (Fig 2B). Under 808 nm illumination, the system selectively projected green light patterns onto the tubes containing fluorophores (Fig 2B-D), correctly identifying their emission intensity while ignoring the non-fluorescent water control. The system allows operators to acquire real-time NIR-II information overlaid directly on the target.

We have also tested solid NIR-II fluorescent materials. An ink mixed with AIE dots (TT3-oCB) was used to print the characters "ZJU BMO-Q" (Fig 2E). Under 808 nm excitation, the system captured a clear NIR-II fluorescence image of the characters (Fig 2F) and projected a corresponding green pattern that precisely overlapped with the printed text (Fig 2G).

Additionally, we conducted experiments using chlorophyll as a natural NIR-II fluorophore, which emits in the NIR-II range under ~690 nm excitation. Monitoring chlorophyll fluorescence has potential agricultural applications, such as assessing tea leaf withering, viral infections, or pesticide distribution. We collected tender, mature, and senescent corn leaves (Fig S1A). Using the system, we clearly observed that both tender and mature leaves emitted detectable fluorescence, whereas the senescent leaves did not (Fig S1B).

### Projection Visibility and Imaging Depth in Biological Tissues

The visibility of projected patterns in a surgical setting can be challenged by tissue coloration and bright ambient light. To evaluate the system's robustness against these factors, we projected a seven-pointed star pattern at maximum luminance onto various tissues of rat, including the skin, fascia, kidney, liver, intestine, and blood, under approximately 2000 lux of ambient illumination. The projected patterns remained clearly discernible, even on darkly pigmented tissues like the liver and blood (Fig 3A).

To assess its performance on fine biological structures, we imaged the ICG-labeled lymphatic system in a mouse leg. The resulting projection onto the intact skin clearly delineated the lymphatic vessels and lymph node, with a measured vessel width of only ~0.2 mm (Fig S2).

To determine the system's imaging depth, we placed centrifuge tubes containing ICG solution under sections of porcine tissue (muscle, belly, and adipose). As shown in Fig 3B, the system successfully localized the tube with a high signal-to-background ratio (SBR) of 4.29, even when it was obscured by 6 mm of muscle tissue. This imaging capability persisted for tissue thicknesses of up to 9 mm (Fig S3). Notably, these results were achieved with a laser excitation power density of only 30 mW/cm², well within safe limits of biological applications.

**Animal Studies: Surgical Navigation for Lymph Node Resection**

Sentinel lymph node biopsy (SLNB) is the standard procedure for staging early-stage breast cancer and relies on accurate lymph node localization(*17*). NIR-II-fluorescence-based projection navigation offers distinct advantages over traditional methods like radiotracers and blue dye(*18*). Radiotracer techniques require specialized hospital infrastructure to manage radiation and are associated with high costs. The blue dye method depends on visual identification, which might fail when the target is obscured by overlying tissue. Furthermore, residual methylene blue can potentially cause tissue necrosis(*19*).

To highlight the practical advantages of our system, we directly compared its performance against the conventional blue dye method. In a rat model, localizing a node marked with methylene blue required a multi-step dissection: the node was invisible through the skin and remained obscured by muscle post-incision, becoming visible only after removal of this overlying tissue (Fig 4A). Conversely, our system instantly localized an ICG-labeled lymph node through the intact skin, providing simple, safe, and reliable real-time guidance before the first cut was made (Fig 4B). This real-time overlay guidance allows surgeons to focus on the resection task. Unlike screen-based navigation, it eliminates the surgeon's frequent visual field switching between the display screen and the surgical site. In addition, it eliminates the need for preoperative skin markings, which can become inaccurate due to tissue displacement after incision. The projection signal remains valid throughout the procedure, localizing deep targets via NIR-II fluorescence and then serving as a continuous and high-contrast "color" guide after the target is exposed.

To further demonstrate the utility of this approach, we performed lymph node resection in rats. In a simulation of SLNB, we injected ICG into a rat's forepaw to label the axillary lymph nodes. As shown in the Fig 4C, the projection clearly revealed the lymphatic vessels and lymph node through the skin. After an incision was made, the projected pattern precisely highlighted the target lymph node against the pink fascia. Guided by the projection, the lymph node was efficiently isolated from the surrounding muscle and excised. This targeted exposure minimizes unnecessary iatrogenic injury, potentially improving patient prognosis. Notably, the size of the projected lymph node on the skin was only 1.5 mm larger than the actual excised node.

We also evaluated the system during a retroperitoneal lymph node resection, a more challenging procedure (Fig 4D). Upon exposing the retroperitoneal cavity, the projection continuously provided high-contrast visualization of the target lymph nodes embedded in adipose tissue.

Throughout the resection, the projection continuously highlighted the nodes, providing unambiguous guidance even as the surgical field shifted. This allowed the operator to maintain focus on the precise dissection while preserving adjacent critical structures, such as blood vessels and the ureter. The successful resection was confirmed by the disappearance of the projected signal at the surgical site. This continuous, intuitive guidance significantly enhanced the convenience and safety of the procedure.

**Validation in Larger Animal Models: Vascular and Lymphatic Visualization**

We then evaluated the applicability of this method in larger animal models, building upon previous clinical validation of ICG-based NIR-II vascular perfusion imaging for orthopedic blood supply assessment. Following abdominal depilation in New Zealand rabbits (Figs 5A), a low-dose intravenous injection of ICG (1 mg/kg via the ear vein) enabled clear visualization of complex subcutaneous vascular networks that are otherwise invisible to the naked eye (Figs 5B and 5C).

The New Zealand rabbit model possesses relatively large hindlimb muscle mass, with popliteal lymph nodes embedded deep within soft tissue—an anatomical configuration that closely mimics clinical scenarios requiring intraoperative identification of obscured lymph nodes. To evaluate lymphatic system visualization, ICG was injected into the hind paw pad (rabbit body weight: 2.5 kg). Pseudocolored NIR-II fluorescence projection imaging enabled clear delineation of lymphatic vessels and the popliteal lymph node from both medial and lateral views (Figs 5D-G). The popliteal node was resolved with a full width at half maximum (FWHM) of 9.31 mm and an SBR of 3.14 (Fig 5H), while two distinct lymphatic vessels were visualized with FWHMs of 2.13 mm and 2.12 mm and corresponding SBRs of 2.46 and 2.33 (Fig 5I). The high spatial resolution and imaging contrast are attributed to reduced tissue scattering in the NIR-II window, resulting from moderated tissue absorption and increased wavelength—factors that collectively enable precise discrimination between target structures and background[20], a level of detail not achievable with conventional NIR-I imaging due to significant scattering interference. Post-euthanasia dissection confirmed the presence of approximately 6.37 mm of overlying muscle tissue above the targeted popliteal lymph node (Fig 5J). Guided by continuous NIR-II fluorescence projection, precise surgical excision of the deeply embedded lymph node was successfully achieved (Fig 5K), demonstrating the system's capability for accurate deep-tissue navigation within anatomically relevant surgical contexts.

**Clinical Feasibility in Diabetic Foot Assessment and Lymphatic System Visualization**

Encouraged by the preclinical results, we proceeded to evaluate the technique's utility for clinical vascular assessment in a patient with diabetic foot ulcers. Following an intravenous injection of ICG, the NIR-II fluorescence imaging and visible light projection system was used. The projection clearly visualized the vascular networks of the lower leg, indicating smooth blood flow without inflammatory reactions (Fig 6A). Quantitative analysis of the original NIR-II fluorescence images revealed a high SBR for the vasculature, reaching 2.80 (Fig 6B). In contrast, the projection over the ulcerated ankle site revealed a diffuse, bright signal at the wound's periphery, which lacked discernible vascular structures (Fig 6C). The NIR-II fluorescence intensity in the wound area was 3.18 times higher than that of the surrounding tissue (Fig 6D). This observation is likely attributable to the enhanced permeability and retention (EPR) effect, where localized inflammation increases vascular leakage, causing ICG to extravasate and accumulate in the interstitial tissue.

We assessed the system's diagnostic capability in a patient with a more severe diabetic foot condition. Preoperative examination of this patient revealed gangrene in the left fourth toe, accompanied by edema and elevated temperature on the dorsal aspect of the foot. After ICG administration, pseudocolored NIR-II fluorescence image projection demonstrated a stark contrast: the gangrenous fourth toe was non-fluorescent, sharply demarcated from the adjacent healthy, fluorescent tissue (Fig 6E). The fluorescence intensity of the normal toes was 6.73 times higher than that of the necrotic toe (Fig 6F), indicating a complete loss of blood supply in the latter. Furthermore, a diffuse fluorescence signal was observed across the entire dorsal foot without clear vascular structure (Fig 6G). This suggested that the inflammation had spread proximally from the affected toe. For comparison, the contralateral right foot of the same patient exhibited clear vascular patterns without signs of inflammation (Fig 6H). Guided by these projection findings, which delineated the boundary between viable and non-viable tissue, the clinical team performed a targeted amputation of the fourth toe and subsequent debridement of the inflamed tissue.

Notably, the excitation power density used in these clinical assessments was 40 mW/cm², well below the American National Standards Institute (ANSI) safety limit for human skin exposure to an 808 nm laser (~329 mW/cm²)(*21*). Additionally, intravenous injection doses were only 0.15 mg/kg, well below daily metabolizable ICG amounts in humans (0.5 mg/kg)(*22*). Therefore, this technology demonstrates a high safety profile. We believe this system, by leveraging the EPR effect to visualize inflammation and directly assessing vascular perfusion, provides an intuitive and clinically valuable tool for determining the extent of ulceration and guiding surgical intervention.

We also evaluated the system's capability to visualize the human lymphatic system. A healthy volunteer received a subcutaneous injection of ICG at the wrist. As shown in Fig 6I, under 808 nm laser excitation, a prominent lymphatic vessel was observed extending from the inner forearm to the elbow. Based on the NIR-II fluorescence image, this lymphatic vessel exhibited a FWHM of approximately 3.79 mm and an SBR of approximately 3.31 (Fig 6J). To confirm that the observed structure was indeed a lymphatic vessel rather than a blood vessel, a commercial vein imaging device (VeinViewer® Flex) was used to visualize the superficial vasculature of the arm. This system projects vascular patterns in real time by exploiting the differential absorption spectra of oxygenated and deoxygenated hemoglobin. The projection results confirmed that the lymphatic vessels and blood vessels followed distinct anatomical paths and orientations (Fig S4). Notably, the lymphatic fluorescence pattern displayed two intensity valleys at the intersection points with blood vessels, suggesting that the lymphatic vessels are located deeper than the blood vessels (Fig 6I). In addition, a relatively clear network of lymphatic vessels was also observed on the outer forearm (Fig S5).

## DISCUSSION

We developed a co-axial NIR-II fluorescence imaging and visible light projection system that renders otherwise invisible NIR-II fluorescence signals directly visible on surgical tissues in real time. This eliminates the need for surgeons to switch visual fields and meets the clinical demand for seamless intraoperative imaging. The system enables accurate in-situ projection of NIR-II fluorescence images with a resolution of 220 μm and a positioning error of less than 0.15 mm. Notably, it maintains high visibility even under bright illumination lighting conditions (up to 2000 lux) and can highlight targets located as deep as 9 mm beneath the tissue surface.

Using this system, we successfully achieved lymphatic and vascular projection, as well as surgical guidance for lymph node resection in both rats and rabbits. The system was capable of visualizing popliteal lymph nodes and lymphatic vessels located beneath approximately 6 mm of muscle tissue on both sides of the rabbit hind limbs. We further evaluated the system's performance in visualizing human vascular and lymphatic networks, capturing detailed structures on the skin. Additionally, the system enabled staging of diabetic foot conditions and assessment of lymphatic vessel morphology, providing valuable reference data for clinical treatment planning.

The system operates well within established safety parameters. The excitation power density of 40 mW/cm² is substantially below the FDA-approved limit of 200 mW/cm² for 808 nm light exposure to human tissue. Potential concerns regarding long-term exposure are mitigated by the system's capability to deliver intermittent, rather than continuous, illumination as needed. The ICG dosing protocols, 0.15 mg/kg intravenously and 0.5 mL, 2.5 mg/mL via subcutaneous injection, remain well within clinically accepted ranges and below the daily metabolizable limits in humans, thereby ensuring minimal risk of systemic exposure.

The current system represents a pivotal step toward seamlessly integrating cutting-edge optical imaging with practical surgical practice. It enables rapid and accurate conveyance of NIR-II fluorescence information while preserving visible-light stereoscopic vision, aligning more closely with surgeons' habitual operating practices and offering a lower learning curve compared to previously reported NIR-II fluorescence-guided surgical navigation techniques. This facilitates smoother surgical workflows and helps reduce the risk of iatrogenic injury. Additionally, the system is more cost-effective and offers higher safety margins compared to ultrasound and X-ray imaging modalities. Its distinct advantages in intraoperative localization of sub-centimeter-deep targets underscore its broad potential for clinical applications.

From an engineering perspective, future efforts may focus on enhancing projection brightness, reducing system complexity, and optimizing image processing algorithms. In addition, two core challenges remain: the development of advanced probes and the expansion of clinical applications. At present, the clinical adoption of NIR-II fluorescence imaging lags significantly behind basic research, primarily due to the limited availability of clinically approved NIR-II fluorescent probes beyond ICG and methylene blue(*23*). They lack the ability to specifically label pathological tissues, such as tumors, which severely constrains the expansion of application scenarios in clinical settings. Meanwhile, current clinical studies are limited by small sample sizes and insufficient exploration of diverse application scenarios, underscoring the need for large-scale but suitable clinical trials to fully validate the clinical value of NIR-II bioimaging technology.

## MATERIALS AND METHODS

### Materials Preparation

Indocyanine Green (ICG) for animal experiments was obtained from Meilunbio (Dalian, China), while clinical-grade ICG was purchased from Dandong Yichuang Pharmaceutical Co., Ltd. and dissolved in the provided sterile water for injection. Polymer dots (Pdots), PbS/CdS quantum dots (QDs), and aggregation-induced emission dots (AIE dots) were synthesized according to previously published protocols (*14-16*). Deionized water was used as the solvent in all experiments, except those involving clinical imaging.

**System Construction**

A custom-designed dual-channel objective was developed to enable co-axial optical paths for NIR-II imaging and visible light projection. The objective incorporated shared lens groups for both visible and NIR-II wavelengths to allow simultaneous focusing, a dichroic prism to ensure optical path co-axiality, and separate lens groups for aberration correction in the visible and NIR-II channels, respectively. During the design process, the objective was treated as a common imaging component for both NIR-II fluorescence and visible-light bright-field dual-channel imaging. It was carefully optimized to achieve identical magnification and to eliminate distortion differences between the two channels.

During NIR-II fluorescence imaging, either an 808 nm or a 690 nm continuous-wave (CW) laser was used as the excitation source. The NIR-II fluorescence signal emitted from biological samples was collected through the custom-designed objective, passed through a 1100 nm long-pass filter (FELH, Thorlabs, USA), and captured by an InGaAs camera (SD640, Tekwin, China). A computer system processed the fluorescence images by performing operations such as intensity adjustment, magnification, and translation, before transmitting the processed images to an LCD panel for real-time display. The LCD had a smaller pixel size than the InGaAs camera, enabling the visualization of fine fluorescence details. To maintain a 1:1 correspondence between the captured fluorescence image and the displayed pattern, the images were resized prior to display. A high-power green LED (electrical power: 60 W) was employed as the projection light source. The green light passed through a set of lens groups and the first polarizer, becoming linearly polarized before illuminating the LCD panel. The polarization direction of the light was modulated by the LCD panel according to the image content. After passing through the second polarizer, these polarization changes were converted into variations in brightness, thereby forming the visible image. Finally, the green light image was projected onto the biological sample through the visible-light optical path of the objective.

**Testing of System's Resolution Limit**

USAF 1951 resolution charts were used as standard test samples. These charts, composed of chrome-coated glass with light-transmitting patterns etched into the chrome layer, contain groups of horizontal and vertical lines with varying sizes. The spatial resolution of the system was evaluated based on the finest set of lines that could be clearly distinguished. The resolution charts were placed directly on the surface of the NIR light source. At a working distance of 28 cm, the NIR-II fluorescence projection system was employed to perform imaging and projection, after which the projected images were frozen. The resolution charts were then replaced with white paper to serve as the projection screen. A high-resolution, visible-light-sensitive camera (Bigeye, Hikvision, China) was used to capture the projected patterns. The clarity of the captured projection images served as an indicator of the system's resolution, which was influenced collectively by factors such as equipment alignment, the performance of the InGaAs camera, the LCD panel resolution, and the optical resolution of the objective lens.

**Characterization of System's Positioning Error**

Checkerboard patterns with edge lengths of 9.57 mm were printed and used for system positioning accuracy testing. First, their bright-field images were captured using a fixed visible-light-sensitive camera. The printed checkerboards were then illuminated with halogen lamps containing NIR-II spectral components, allowing the NIR-II fluorescence projection system to image the patterns and project corresponding green pseudocolored checkerboards back onto the surface. After freezing the projected images, the printed checkerboards were replaced with white paper to serve as a projection screen. The visible-light-sensitive camera then captured the projected checkerboard patterns on the paper. The positioning accuracy of the system was evaluated by comparing the corner points of the original printed checkerboards and those of the projected patterns. The visible-light-sensitive cameras used had RGBW pixel arrangements, and the W (white) channels were extracted for analysis. Corner point detection was performed using MATLAB, and the pixel-wise distances between corresponding corner point pairs in the original and projected images were calculated. Knowing the physical edge length of the checkerboard (9.57 mm), the pixel distances were converted into millimeters. Given that surgical targets often have irregular, undulating surfaces, the NIR-II fluorescence projection system is required to maintain high positioning accuracy within its depth of field. Therefore, after fixing the system position, the checkerboards were repositioned at various depths within the depth of field to evaluate positioning errors under different spatial conditions.

**Animal Handling**

All animals were obtained from the Zhejiang University School of Medicine. The experimental subjects included 20 g female ICR mice, 300 g female Sprague-Dawley (SD) rats, and 2 kg New Zealand white rabbits. All procedures were conducted in accordance with the guidelines approved by the Laboratory Animal Ethics Committee of Zhejiang University (approval number: ZJU20220283).

To evaluate the projection visibility on various biological tissues, the NIR-II fluorescence projection system was configured to project seven-pointed stars at maximum brightness. Under indoor lighting conditions of approximately 2000 lux, the projection patterns were directed onto various colored tissues of rats, including the abdominal skin, fascia, kidney, liver, intestine, and blood. The resulting images were captured using mobile phones. For imaging the leg lymphatic system and retroperitoneal lymph nodes of mice, 100 μL of ICG aqueous solution (1 mg/mL) was subcutaneously injected into the hind paw pads. The ICG subsequently traveled through the lymphatic vessels, sequentially staining the popliteal and retroperitoneal lymph nodes. Similarly, to visualize the retroperitoneal and axillary lymph nodes in rats, 500 μL of ICG solution (1 mg/mL) was injected into the hind and forepaw pads, respectively. For vascular imaging in rabbits, 2 mL of ICG aqueous solution (1 mg/mL) was intravenously injected into the ear vein after the animal was anesthetized. To image and project the rabbit's leg lymphatic system, an equal volume of ICG solution was injected into the paw pads. Prior to all experiments, the imaging sites of all animals were depilated using depilatory cream to reduce interference from fur.

**Patient Recruitment and Clinical Data Collection**
Volunteers were recruited from patients undergoing treatment for diabetic foot and lymphedema at the affiliated hospital of Zhejiang University. After informed and fully understanding the study procedures, the volunteers participated in clinical trials utilizing our NIR-II fluorescence projection system. The study was approved and supervised by the Ethics Committee of the Sir Run Run Shaw Hospital (approval numbers: 20220279 and 20250275). Patients with diabetic foot received intravenous injections of ICG (3 mL, 2.5 mg/mL). In addition, a healthy volunteer was recruited and received a single-point subcutaneous injection of ICG at the wrist, followed by NIR-II imaging and projection for lymphatic visualization of the arm and wrist.

**List of Supplementary Materials**

Fig S1 to S5

**Acknowledgments:**
**Funding:**
    National Key R&D Program of China (2024YFF1206700)
    National Natural Science Foundation of China (U23A20487)
    Dr. Li Dak Sum & Yip Yio Chin Development Fund for Regenerative Medicine, Zhejiang University, China
    Hangzhou Chengxi Sci-tech Innovation Corridor Management Committee funded project
    China Postdoctoral Science Foundation funded project (BX20220260 and 2023M733038)

**Author contributions:**
    Conceptualization: JQ, HL, YZ
    Methodology: YZ, XL, ZF, JQ
    Investigation: YZ, XL, ZL, CL, JY, JF, SS
    Visualization: YZ, XL, ZF
    Supervision: JQ
    Writing—original draft: YZ, XL, ZL
    Writing—review & editing: JQ

**Competing interests:** Authors declare that they have no competing interests.

**Data and materials availability:** All data are available in the main text or the supplementary materials.

**Figures**

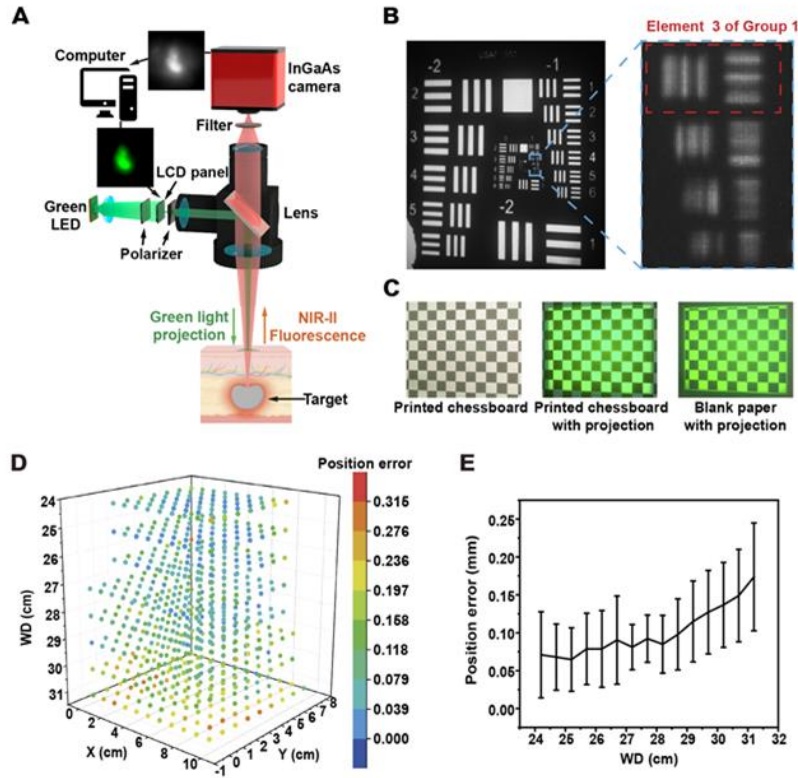

**Fig. 1. Design and performance characterization of the co-axial NIR-II fluorescence imaging and visible light projection system.** (A) Schematic diagram illustrating the system architecture. (B) A projected USAF 1951 test pattern demonstrates that the smallest resolvable feature (Group 1, Element 3, dashed box) is 2.55 lp/mm, corresponding to a spatial resolution of the system of approximately 220 μm. (C) Methodology for assessing positioning accuracy. Images show: (i) the original printed checkerboard used as a reference, (ii) the projected pattern overlaid on the checkerboard, and (iii) the projected pattern displayed on a white paper screen for error calculation. Accuracy is quantified by measuring the displacement between corresponding corner points of the printed and projected patterns. (D) A spatial map of co-localization error across the depth of field, with working distances ranging from 24 to 31 cm. Color coding indicates the magnitude of displacement error. (E) Quantitative analysis of co-localization error as a function of working distance. The system maintains sub-millimeter accuracy throughout the 7 cm depth of field, with a minimal error of less than 0.15 mm at the optimal working distance of 28 cm.

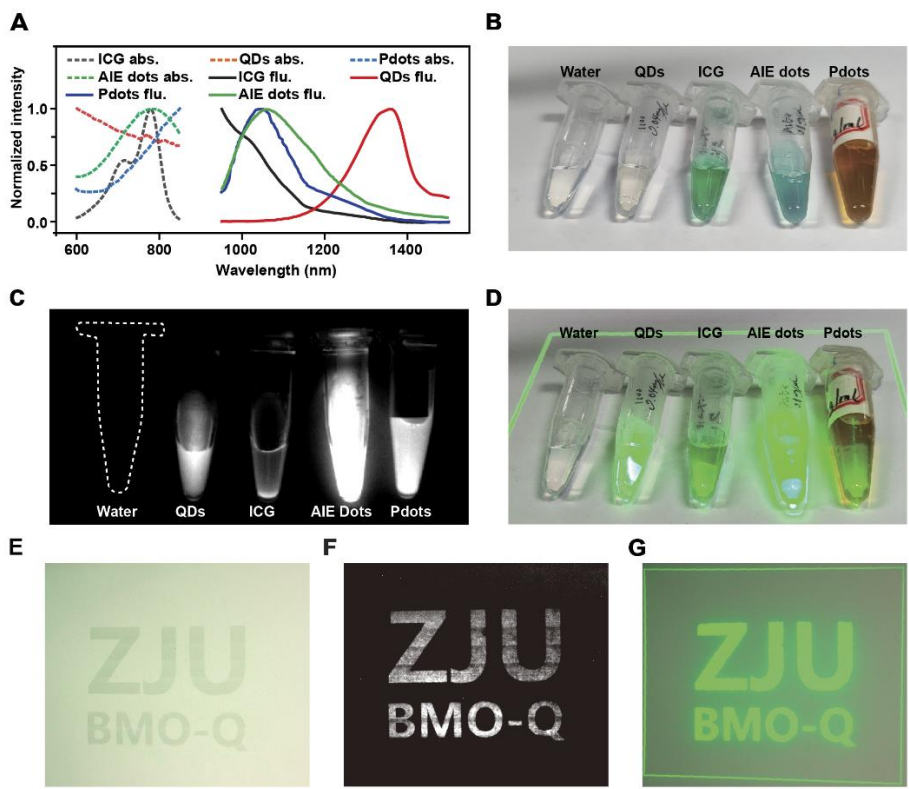

**Fig. 2. In vitro positioning tests utilizing NIR-II fluorescent materials.** (A) Absorption and fluorescence spectra of aqueous dispersions of various NIR-II fluorescent materials, including ICG, PbS/CdS QDs, Pdots (L1057), and AIE dots (TT3-oCB). (B) Bright-field image of aqueous dispersions of ICG, PbS/CdS QDs, Pdots (L1057), AIE dots (TT3-oCB), and water (used as a control). (C) NIR-II fluorescence image of the same dispersions under 808 nm laser excitation. All fluorescent materials exhibit strong NIR-II emission, whereas water shows no fluorescence. (D) Bright-field image overlaid with projected green pseudocolored NIR-II fluorescence, facilitating clear visual differentiation of the various fluorophores. (E) Bright-field image of a paper printed with the text "ZJU BMO-Q" using black ink blended with AIE dots (TT3-oCB). (F) NIR-II fluorescence image of the printed pattern under 808 nm laser excitation. (G) Bright-field image with green pseudocolored fluorescence projection, showing a precise alignment between the projected pattern and the printed NIR-II fluorescent ink.

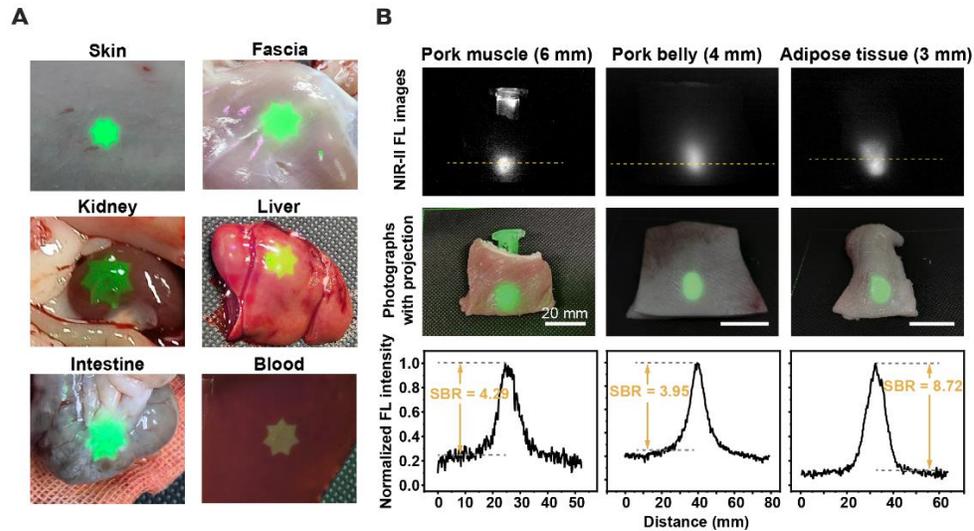

**Fig 3. Projection visibility on biological tissues and deep-tissue target detection.** (A) Projected star-shaped patterns onto various rat tissues, including skin, fascia, kidney, liver, intestine, and blood, under bright ambient illumination (2000 lux). The projections remain clearly visible despite differences in tissue color and texture. (B) Evaluation of deep-tissue target detection using coverings composed of tissues with varying types and thicknesses. The target consists of a 0.1 mg/mL solution of ICG in a microcentrifuge tube. Tissue overlays include porcine muscle, pork belly, and porcine adipose tissue. Top row: NIR-II fluorescence images showing the target beneath different tissues. Middle row: Bright-field images with overlaid projection patterns. Bottom row: Fluorescence intensity profiles along the yellow dashed lines shown in the top row, illustrating corresponding SBRs. The excitation power density was maintained below 40 mW/cm² in all experiments.

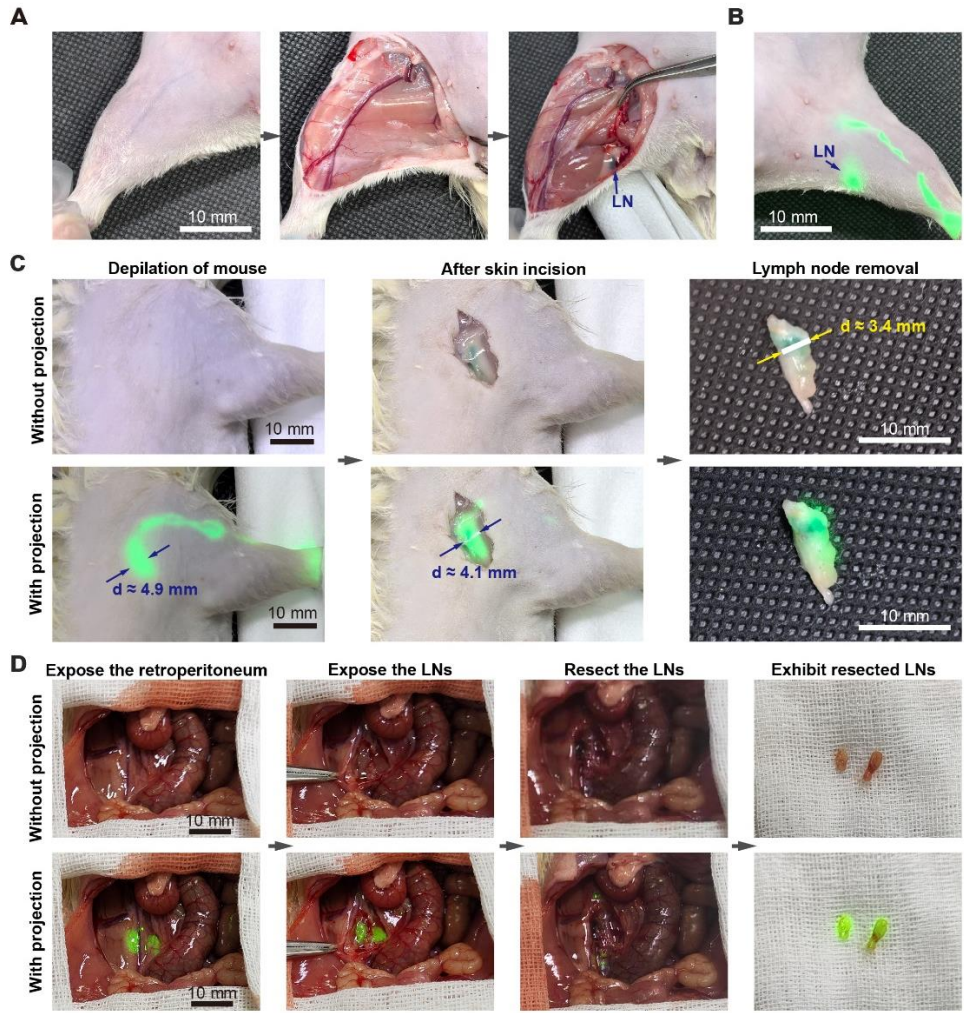

**Fig 4. Image-guided lymph node resection in rats using the co-axial NIR-II fluorescence imaging and visible light projection system.** (A) Bright-field images of the right hindlimb of a rat pre-injected with methylene blue. After skin removal, the methylene blue-stained lymph node remains visually undetectable due to overlying muscle tissue. Only following dissection of the muscle, the dark blue-stained lymph node (indicated by the arrow) becomes directly visible. (B) Bright-field image of the contralateral (left) popliteal region pre-injected with ICG. The lymph node (arrow) and associated lymphatic vessels are clearly visualized through intact skin via the green projected pattern from the system. (C) ICG-pre-injected forepaw of the rat before and after skin incision. The projected fluorescence pattern enables accurate transcutaneous localization of the lymph node prior to incision and continues to provide visual guidance post-incision. Excised ICG-stained lymph node showing distinct green fluorescence ex vivo. (D) Resection of retroperitoneal lymph nodes: Exposed retroperitoneal region with visible projection marking two lymph nodes embedded in adipose tissue; Direct visualization of the nodes following exposure, with projection highlighting the targets; Post-resection view showing no residual projection signal at the surgical site; The excised lymph nodes. Excitation power density was maintained at 20 mW/cm² throughout all procedures.

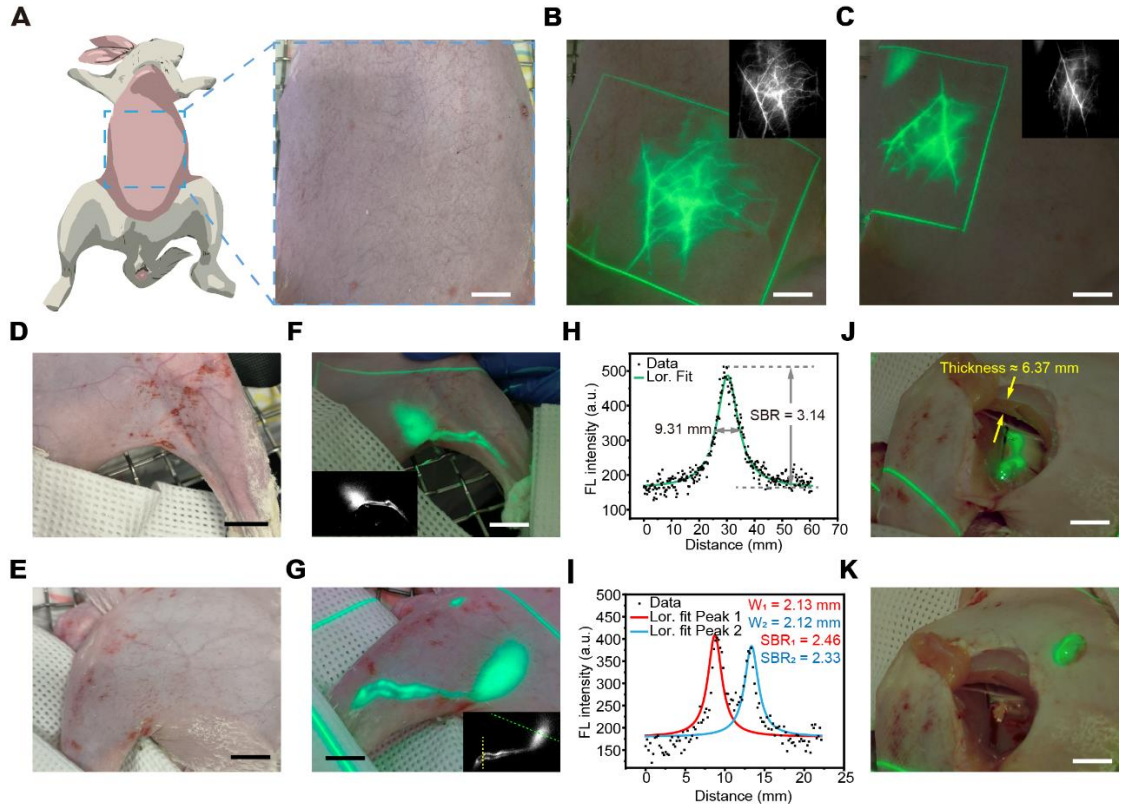

**Fig 5. In vivo vascular and lymphatic imaging in rabbits using the co-axial NIR-II fluorescence imaging and visible light projection system.** (A) Schematic illustration showing the removal of abdominal hair from the rabbit, and the bright-field image of the rabbit's abdominal region. (B, C) Projected pseudocolored NIR-II fluorescence images of abdominal vasculature acquired after intravenous injection of ICG via the ear vein, under 808 nm laser excitation. (Insets: corresponding raw NIR-II fluorescence images.) (D, E) Bright-field images of the medial and lateral aspects of the hindlimb from a separate rabbit with ICG-labeled lymphatic system. (F, G) Bright-field images overlaid with projected green pseudocolored NIR-II fluorescence images of the lateral (F) and medial (G) hindlimb, clearly delineating the lymph node and associated lymphatic vessels. (Insets: corresponding raw NIR-II fluorescence images.) (H, I) Fluorescence intensity profiles extracted along the green and yellow dashed lines in (G), respectively: (H) Profile across the lymph node (full width: 9.31 mm; SBR = 3.14). (I) Profile across two lymphatic vessels (widths: 2.13 mm and 2.12 mm; SBRs = 2.46 and 2.33, respectively). (J) Projection-guided muscle dissection for lymph node exposure. The muscle covering the lymph nodes was measured to be 6.37 mm thick. (K) The excised lymph node. Excitation power density was maintained at 20 mW/cm² throughout. Scale bar: 20 mm.

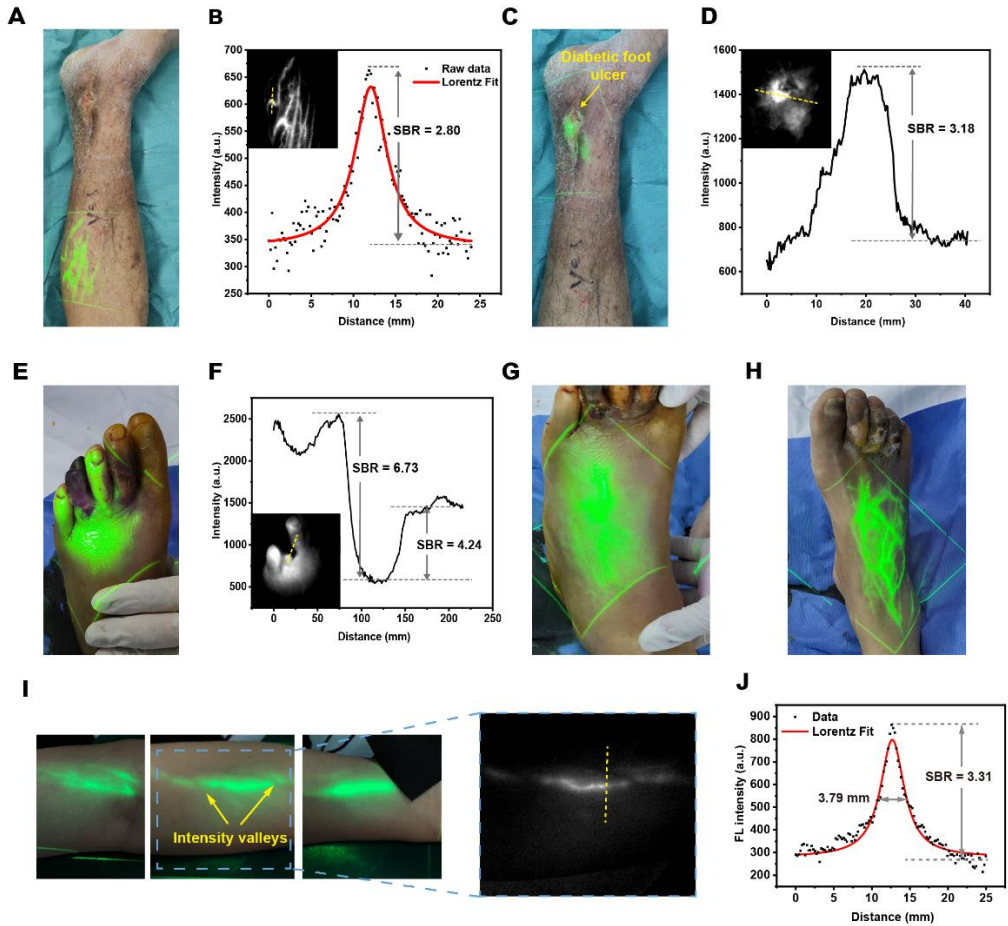

**Fig 6. Clinical translation of the co-axial NIR-II fluorescence imaging and visible light projection system.** (A) Bright-field image of the calf in Patient 1 (diabetic foot) following intravenous injection of ICG, overlaid with the projected green pseudocolored NIR-II fluorescence image. The vascular network is clearly visualized. (B) Fluorescence intensity profile along the dashed line in the inset, showing an SBR of 2.80 for the vessel (Inset: corresponding NIR-II fluorescence image of (A)). (C) Bright-field image of the ankle region in Patient 1, overlaid with the projected green pseudocolored NIR-II fluorescence image. A bright, ring-like pattern surrounding an ulcer indicates inflammation, which is accompanied by a loss of distinct vascular structures. (D) Fluorescence intensity profile along the dashed line in the inset. The fluorescence intensity within the inflamed region is approximately 3.18-fold higher than that of the adjacent tissue (Inset: corresponding NIR-II fluorescence image of (C)). (E) Bright-field image of the left toes in Patient 2 (more severe diabetic foot), overlaid with the projected green pseudocolored NIR-II fluorescence image. The patient presents with gangrene in the fourth toe and inflammatory dorsal swelling. An absence of fluorescence signal in the gangrenous fourth toe contrasts with the strong signal in adjacent toes and the dorsal foot. (F) Fluorescence intensity profile along the dashed line in the inset, confirming a marked decrease in fluorescence intensity in the necrotic toe compared to the surrounding viable tissue (an intensity difference of approximately 6.73-fold, Inset: corresponding NIR-II fluorescence image of (E)). (G, H) Bright-field images of the left (G) and right (H) dorsal feet of Patient 2, overlaid with the projected pseudocolored fluorescence images. The vascular structure in the pathologically affected left foot (G) appears indistinct, whereas it is comparatively clear in the contralateral right foot (H). Excitation power density: 40 mW/cm². (I) Bright-field images of lymphatic vessels in the forearm of a healthy volunteer, overlaid with the projected green pseudocolored NIR-II fluorescence images acquired 4 hours after subcutaneous injection of ICG at the wrist. (Right inset: corresponding NIR-II fluorescence image). (J) Fluorescence intensity profile along the dashed line in (I). The lymphatic vessel exhibits an SBR of 3.31 and a width of 3.79 mm. Excitation power density: 20 mW/cm².

# Supplementary Materials for

**NIR-II Fluorescence Project Technology for Augmented Reality Surgical Navigation**

Yuhuang Zhang *et al.*

*Corresponding author. Email: qianjun@zju.edu.cn

**This PDF file includes:**

Figs. S1 to S5

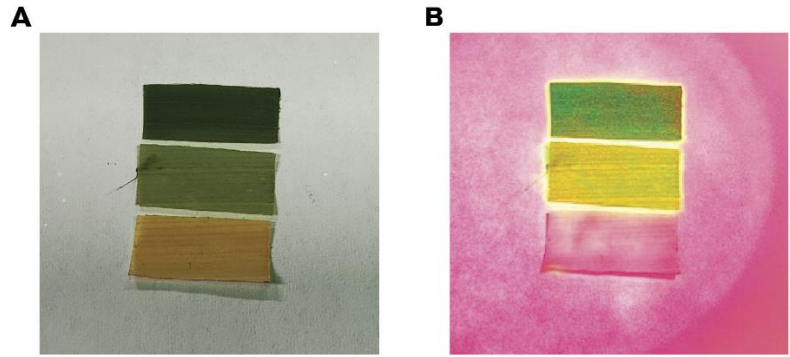

**Fig S1. Demonstration of chlorophyll NIR-II fluorescence detection in maize leaves using the projection system.** (A) Photographs of maize leaves representing distinct developmental stages: emerging (top), mature (middle), and senescent (bottom). (B) Fluorescence imaging and projection under ~690 nm excitation using the developed system. The tender green and mature leaves exhibit clear NIR-II fluorescence, visualized here as projected green light. In contrast, the yellowed leaf, exhibiting significantly reduced chlorophyll content, shows negligible fluorescence and appears like the background.

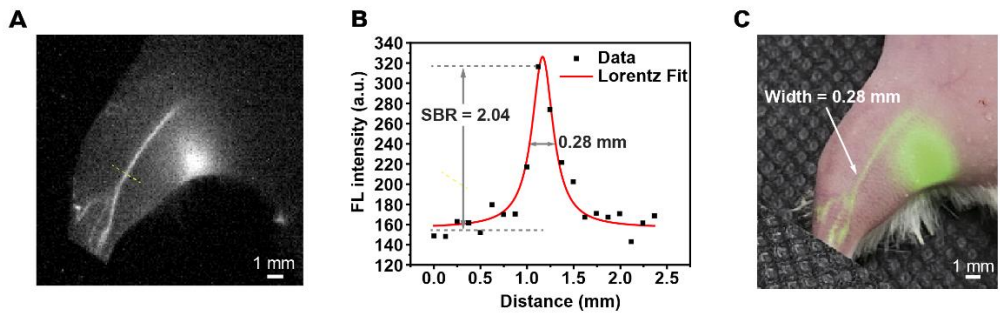

**Fig S2. In vivo mapping of lymphatic system in mouse using the projection system.** (A) A NIR-II fluorescence image of ICG-labeled lymph node and lymphatic vessels. (B) Intensity profile along yellow dashed line in (A). Gaussian fitting reveals a lymphatic vessel width (FWHM) of 0.28 mm with SBR = 2.04. (C) The projected pseudocolored fluorescence image demonstrating precise co-localization of lymph node and submillimeter lymphatic vessels (0.28 mm width).

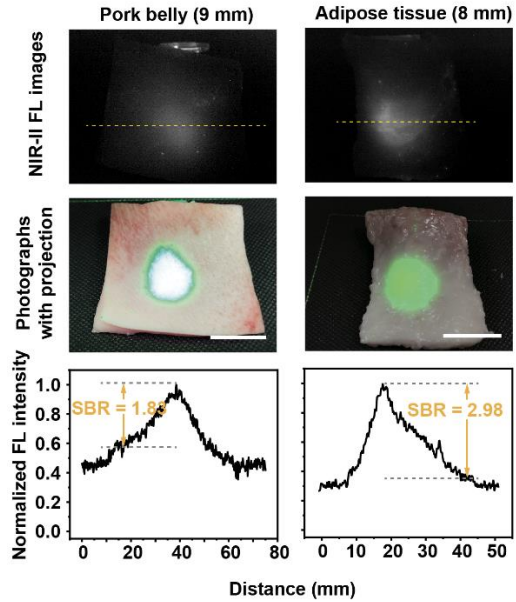

**Fig S3. Porcine pork belly (9 mm) and adipose tissue (8 mm) were overlaid on microcentrifuge tubes containing 0.1 mg/mL ICG solution.** The NIR-II fluorescence projection system maintained fluorescence localization capability at these depths with sustained SBRs of 1.83 and 2.98 respectively, as well as precise projection.

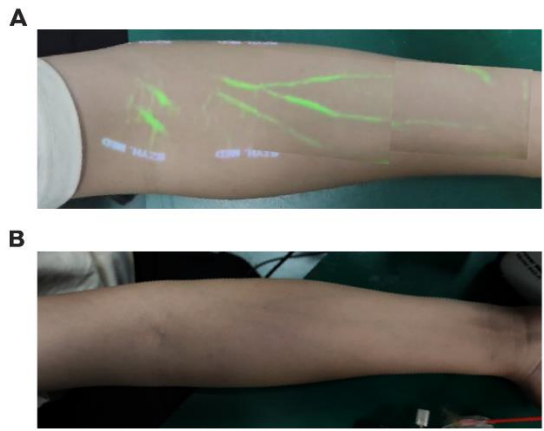

**Fig S4. Comparison of vascular and lymphatic visualization in the human forearm.** (A) Projection of superficial venous structures on the right forearm of a healthy volunteer using a commercial vein imaging system, revealing a structure distinct from the lymphatic vessels shown in Fig 6I. (B) Bright-field image of the same forearm. Note: The limited field of view (FOV) of the commercial imaging system necessitated image stitching to generate the complete projection.

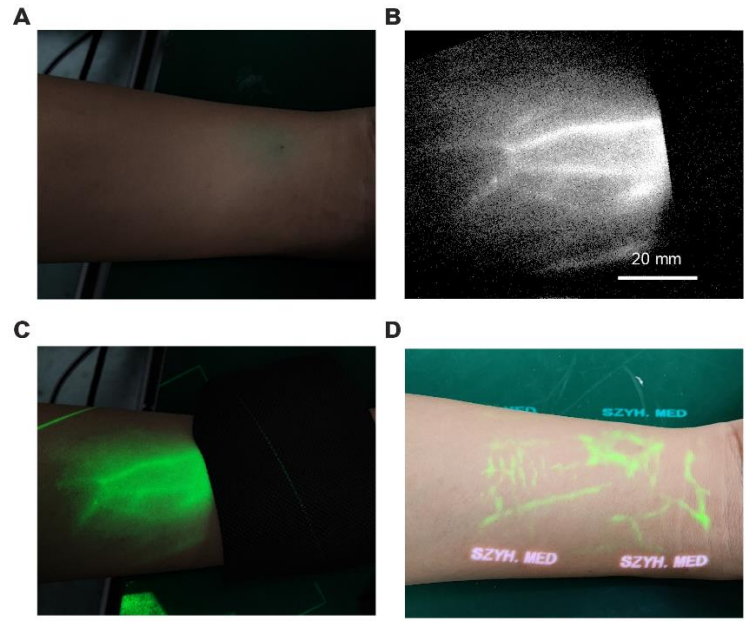

**Fig S5. Lymphatic mapping following subcutaneous injection of ICG.** (a) Bright-field image of the injection site at the wrist. (b) NIR-II fluorescence image showing the lymphatic vasculature (injection site masked to prevent signal oversaturation). (c) Projected pseudocolored NIR-II fluorescence image illustrating lymphatic channels. (d) Comparative projection using a commercial vein imaging system (VeinViewer® Flex), confirming distinct spatial patterns between the vascular (d) and lymphatic (c) systems. Imaging was performed 4 hours after ICG injection (0.5 mL, 0.1 mg/mL).